\documentclass[journal,]{IEEEtran}
\usepackage{amsmath,graphicx,amssymb}
\usepackage{bm}
\usepackage{tikz}
\usepackage{pgfplots}
\usepackage{csvsimple}
\usepackage{tumcolor}
\usepackage{amsmath}
\usepackage{amsthm,amssymb}

\usepackage[noadjust]{cite}
\usepackage{caption}
\usepackage{subcaption}
 
\DeclareMathAlphabet{\mathbit}{OML}{cmr}{bx}{it}

\DeclareMathOperator*{\argmax}{argmax}

\DeclareMathOperator{\Hermitian}{H}
\newcommand{\He}{{\Hermitian}}

\DeclareMathOperator{\diag}{diag}

\DeclareMathOperator{\Exp}{E}

\DeclareMathOperator{\modulo}{mod}

\newcommand{\bth}{{\bm{\theta}}}
\newcommand{\supk}{{(k)}}

\definecolor{MyPurple}{rgb}{0.4,0,0.8}
\definecolor{MyGreen}{rgb}{0,0.5,0}
\definecolor{MyBrown}{rgb}{0.5,0.3,0.1}
\definecolor{MyCyan}{rgb}{0.2,1,1}
\definecolor{MyPink}{rgb}{1,0.2,1}

\tikzset{DashPattern/.style={dash pattern=on 3pt off 2pt on 7pt off 2pt}}
\newcommand{\lineWidth}{3pt}
\tikzset{PlotGenieML/.style={mark options={solid},color=black, line width=\lineWidth}}
\tikzset{PlotDML/.style={mark options={solid},color=MyPink, line width=\lineWidth,mark=|,mark size=5pt}}
\tikzset{PlotSML/.style={mark options={solid},color=TUMLightBlue, line width=\lineWidth,mark=*}}
\tikzset{PlotGLS/.style={mark options={solid},color=TUMBeamerYellow, line width=\lineWidth,mark=square*}}
\tikzset{PlotSPICE/.style={mark options={solid},color=TUMBeamerDarkRed, line width=\lineWidth,mark=diamond*,mark size=4pt}}
\tikzset{PlotSPICEiter/.style={mark options={solid},color=TUMBeamerOrange, line width=\lineWidth,mark=star,mark size=5pt}}
\tikzset{PlotOMP/.style={mark options={solid},color=MyBrown, line width=\lineWidth,mark=o,mark size=4pt}}
\tikzset{PlotCovOMP/.style={mark options={solid},color=MyPurple, line width=\lineWidth,mark=pentagon*,mark size=4pt}}

\tikzset{PlotMCENet/.style={mark options={solid},color=TUMBlue, line width=\lineWidth,dash pattern= on 8pt off 3pt,mark=x,mark size=5pt}}

\tikzset{PlotMUSICAffInv/.style={mark options={solid},color=TUMBeamerOrange, line width=\lineWidth,dashed,mark=*}}
\tikzset{PlotMUSICLogFro/.style={mark options={solid},color=TUMBeamerRed, line width=\lineWidth,dashed,mark=square*}}
\tikzset{PlotMUSICJBD/.style={mark options={solid},color=MyGreen, line width=\lineWidth,dashed,mark=diamond*,mark size=4pt}}
\tikzset{PlotMUSIC/.style={mark options={solid},color=TUMDarkerBlue, line width=\lineWidth,dashed,mark=star,mark size=5pt}}
\tikzset{PlotMUSICFro/.style={mark options={solid},color=TUMLightBlue, line width=\lineWidth,dashed,mark=triangle*,mark size=4pt}}
\tikzset{PlotMUSICChol/.style={mark options={solid},color=MyBrown, line width=\lineWidth,dashed,mark=pentagon*,mark size=4pt}}
\tikzset{PlotMUSICJDiv/.style={mark options={solid},color=MyPurple, line width=\lineWidth,dashed,mark=|,mark size=5pt}}

\tikzset{PlotClaPeak/.style={mark options={solid},color=MyGreen, line width=\lineWidth,DashPattern,mark=square*}}
\tikzset{PlotClaPeakSM/.style={mark options={solid},color=MyBrown, line width=\lineWidth,DashPattern,mark=diamond*,mark size=4pt}}
\tikzset{PlotClaMulti/.style={mark options={solid},color=TUMBeamerYellow, line width=\lineWidth,DashPattern,mark=star,mark size=5pt}}
\tikzset{PlotClaChain/.style={mark options={solid},color=TUMLightBlue, line width=\lineWidth,DashPattern,mark=triangle*,mark size=4pt}}
\tikzset{PlotClaChainProj/.style={mark options={solid},color=TUMBlue, line width=\lineWidth,DashPattern,mark=pentagon*,mark size=4pt}}
\tikzset{PlotClaChainProjWO/.style={mark options={solid},color=TUMBeamerRed, line width=\lineWidth,DashPattern,mark=|,mark size=5pt}}
\tikzset{PlotClaChainEst/.style={mark options={solid},color=TUMBeamerOrange, line width=\lineWidth,DashPattern,mark=x,mark size=5pt}}
\tikzset{PlotClaChainProjEst/.style={mark options={solid},color=MyPurple, line width=\lineWidth,DashPattern,mark=o,mark size=4pt}}
\tikzset{PlotClaChainProjWOEst/.style={mark options={solid},color=TUMBeamerGreen, line width=\lineWidth,DashPattern,mark=*}}

\pgfplotsset{tick label style={font=\small}}

\title{ChainNet: Neural Network-Based Successive Spectral Analysis}
\author{Andreas~Barthelme and Wolfgang~Utschick,~\IEEEmembership{Fellow,~IEEE}\\Professur f\"ur Methoden der Signalverarbeitung, Technical University Munich, 80290 Munich, Germany\\Email: \{a.barthelme, utschick\}@tum.de}

\begin{document}
	%\ninept
	%
	\maketitle
	\begin{abstract}
		We discuss a new neural network-based direction of arrival estimation scheme that tackles the estimation task as a multidimensional classification problem. The proposed estimator uses a classification chain with as many stages as the number of sources. Each stage is a multiclass classification network that estimates the position of one of the sources. This approach can be interpreted as the approximation of a successive evaluation of the maximum a posteriori estimator. By means of simulations for fully sampled antenna arrays and systems with subarray sampling, we show that it is able to outperform existing estimation techniques in terms of accuracy, while maintaining a very low computational complexity.
	\end{abstract}

	\section{Introduction}
	Direction of Arrival (DoA) estimation denotes the process of determining the directions from which electromagnetic waves are impinging onto an antenna array. This task plays a crucial part in the localization of electromagnetic sources and reflectors, which is relevant in military and civil applications such as autonomous driving, aviation, astronomy, and mobile communications.
	
	Traditionally, in these areas, suitable stochastic models for the received signals at the individual antenna elements are available. From these models, a plethora of powerful algorithms for DoA estimation have been derived in the past, cf. \cite{Krim1996,Trees2002}. These DoA estimation strategies encompass beamformer, super resolution and maximum likelihood (ML) techniques. Yet, there are still some applications, where existing solutions do not provide a satisfying performance. For example, systems with subarray sampling, where fewer radio frequency (RF) chains than antenna elements are used to sequentially sample the antenna array, belong to a class of underdetermined systems, where the classical approaches lead to an unsatisfactory performance \cite{Barthelme2020b}. Therefore, the capabilities of data-based neural network (NN) approaches for DoA estimation have been studied recently. These methods may not only yield a better accuracy in complicated scenarios, but are in general computationally attractive.
	
	Existing NN approaches for DoA estimation fall into three different categories \cite{Barthelme2020b}. The first category of NNs poses the DoA estimation problem as a classification task, cf. \cite{Chakrabarty2017,Liu2018,Chakrabarty2019,Ozanich2020,Yao2020,Papageorgiou2020a}. For this purpose, the field of view is divided into several non-overlapping sectors. Rather than estimating DoAs directly, the NN should determine which of these sectors contains an active source. For the single source case, this problem is a simple multiclass classification problem, where the number of sectors corresponds to the number of classes. However, for multiple sources, finding a suitable model becomes much more difficult.
	
	For the second category, the goal of the NN is to estimate a discretized spatial spectrum, from which we can derive the DoAs. One possible spatial spectrum that can be utilized as such a proxy is the transmit power spectrum \cite{Wu2019}. 
	Alternatively, in \cite{Elbir2020}, the MUSIC spectrum that corresponds to the observed received signals is used as the target of a regression network. Similarly, a pseudo-spectrum is utilized in \cite{Izacard2019} for frequency estimation in a multisinusoidal signal.
	
	A more direct approach is the rationale behind methods from the third category. There, the idea is to produce the DoA estimates at the output of the NN. Then, the cost function of interest, e.g., MSE, can be directly used for the training of these regression networks, which works without a discretization of the field of view. In \cite{Guo2020}, two different DoA regression networks are used to resolve two narrowly spaced sources. The choice between the two networks is realized by an SNR classification network. A more general approach is presented in \cite{Bialer2019}, where a NN is proposed that simultaneously estimates the number of sources and their respective DoAs.
	
	Recently, NN-based DoA estimators for systems with subarray sampling have been studied in \cite{Barthelme2020b,Barthelme2021}. The method presented in \cite{Barthelme2020b} falls in the third category. There, a regression-based NN for this scenario has been proposed that trains the network on the mean circular error cost function. In contrast, the estimator discussed in \cite{Barthelme2021} follows a rationale that has been derived particularly for the subarray sampling framework. It estimates the covariance matrix of the fully sampled array from the subarray sample covariance information, and then, applies a MUSIC estimator \cite{Schmidt1986} to obtain the DoA estimates.
	
	In this work, we propose a new estimation scheme, which we refer to by ChainNet, that follows the rationale of the first category of algorithms. The idea is to construct a chain of multiclass classifiers such that each stage provides a DoA estimate for one of the sources. Thereby, the estimation is performed successively, i.e. the input data in each node does not only consist of the observations, but also includes the estimates from previous stages of the chain. The proposed NN-based estimator exhibits an insightful theoretical interpretation, as it can be viewed as an approximation to a successive implementation of the theoretical optimal MAP estimator. In addition to the attractive theoretical parallels to the MAP estimator, we show in simulations that the ChainNet approach is able to significantly outperform other classical and NN-based estimators in terms of achievable accuracy.
	
	\section{System Model}
	\label{sec:sysmodel}
	We consider systems with subarray sampling and fully sampled antenna arrays with $M$ antenna elements. Systems with subarray sampling only use $W<M$ RF chains and a switching network to sample the received signals at the $M$ antenna elements consecutively. Hence, the received signals at all $M$ antennas are not sampled simultaneously. Instead, in one time slot, the received signals of $W$ antenna elements that together form one subarray are captured. In total, the switching network selects $K$ different subarrays, for which we collect $N$ snapshots each. Under the assumption that $L$ sources located in the far-field of the antenna array are transmitting narrow-band signals, the received signal $\bm{y}^{(k)}(n)$ for the $k$-th subarray in the $n$-th snapshot reads as
	\begin{equation}
	\bm{y}^{(k)}(n) = \bm{G}^{(k)}\left(\bm{A}(\bth)\bm{s}^{(k)}(n)+\bm{\eta}^{(k)}(n)\right),\label{eq:sysmodel}
	\end{equation} 
	where the array manifold $\bm{A}(\bm{\theta})\in\mathbb{C}^{M\times L}$ gathers the response of the whole array on the DoAs $\bth=[\theta_1,\dots,\theta_L]$, $\bm{G}^{(k)}\in\{0,1\}^{W\times M}$ selects the active antenna elements that form the $k$-th subarray, and $\bm{\eta}^{(k)}\sim\mathcal{CN}(\mathbf{0},\sigma_\eta^2\mathbf{I}_M)$ is some i.i.d.\ additive white Gaussian noise. Furthermore, the transmit signals $\bm{s}^{(k)}$ are assumed to be temporally uncorrelated, complex Gaussian distributed, i.e., $\bm{s}^{(k)}(n)\sim\mathcal{CN}(\mathbf{0},\bm{C}_{\bm{s}})$ and $\text{E}[\bm{s}^{(k)}(n)\bm{s}^{\He,(q)}(m)]=\mathbf{0}, \forall n\neq m, \forall k\neq q$. For the resulting distribution of the received signals, which is given by
	\begin{equation}
	\bm{y}^{(k)}(n)\sim\mathcal{CN}\left(\mathbf{0},\bm{G}^{(k)}\bm{A}(\bth)\bm{C}_{\bm{s}}\bm{A}^\He(\bth)\bm{G}^{\He,(k)}+\sigma_\eta^2\mathbf{I}_W\right),
	\label{eq:smldist}
	\end{equation}
	sufficient identifiability conditions even for $L\geq W$ exist if the transmit signals are spatially uncorrelated, i.e., $\bm{C}_{\bm{s}}$ is diagonal \cite{Suleiman2018}.
	
	The system model for the fully sampled case is obtained from (\ref{eq:sysmodel}) for $W=M$ and $K=1$ and follows the same assumptions on the densities of the noise and transmit signals.
	
	\section{DoA Estimation as a Classification Task}
	
	In this section, we discuss several schemes that formulate the DoA estimation problem as a classification task. All of the presented methods share a common notion. The field of view from $[0,2\pi)$ is split into $G$ distinct sectors with the same width $\theta_{\text{w}}=2\pi/G$, such that the $g$-th sector ranges from $(g-1)\theta_{\text{w}}$ to $g\theta_{\text{w}})$. Firstly, we are  not interested in obtaining the continuous estimates of the DoAs directly, but in the indices of the sectors that contain an active source.\footnote{If the NN determines that the $g$-th sector contains an active source, we can obtain an estimate for the DoA of this source by taking the mid point of this sector $(g-\frac{1}{2})\theta_{\text{w}}$ or by using further refinement techniques{, e.g., quadratic interpolation of the classifier outputs}.} {Thus, the training dataset consists of labeled tuples $ (\bm{y}_n,g_n(1),\dots,g_n(L)) $, where $ \bm{y}_n $ denotes a placeholder for the respective input vector of the classifier and $ g_n(\ell) \in \{1,\dots,G\}$ denotes the sector containing the true $\ell$-th DoA of the $n$-th training sample.} 
	
	\subsection{Single Source Case}
	\label{sec:singlesource}
	We first consider the single source case. For this scenario, the formulation of the classification problem is straight-forward and has been presented in \cite{Chakrabarty2017,Li2018}. Since there is only one source, {$ L = 1$}, we are trying to find the one sector {$ g $} out of all $G$ sectors, which is most probable to include the active source. This is a standard multiclass classification problem, which can be solved by training on the so called categorical cross-entropy loss function. To that end, we encode the index of the true sector for each training data sample in a one-hot vector, i.e., the label of the true sector is $1$, whereas the label for all other sectors is $0$. At each neuron of the output layer a softmax operation is applied, which produces $G$ outputs {$z_g(\bm y;\bm{w})$}, $g = 1,\dots,G $, {whoses values} are between zero and one and their sum is the value one \cite{Bridle1989}. In combination with a training based on the cross-entropy loss, given by 
	\begin{equation}
	{\max_{\bm{w}}\log\left(z_{g_n}(\bm{y_n};\bm{w})\right),}
	\label{eq:crossentropy}
	\end{equation}
	where $\bm{w}$ are the weights and biases of the NN and {$g_n$} is the index of the active sector of the {$n$-th input vector} {$\bm{y}_n$}, these output values {$z_{g}(\bm{y};\bm{w})$} can be interpreted as estimates of the posterior probabilities {$ p_{g|\bm{y}} \approx z_{g}(\bm{y})$ } for each class $g$ conditioned on $\bm{y}$.
	As has been discussed in \cite{Lecun1998}, the training based on (\ref{eq:crossentropy}) can be interpreted as a heuristic approach to the theoretically optimal MAP estimator, as the NN is tuned to maximize the estimate of the posterior probability of {the correct sector label of the input vector.}
	
	\subsection{Multiple Source Case}
	For more than one source the classification task gets more difficult. First of all, let us assume that in each sector, there can be at most one source.  
	Note that this assumption requires $G$ to be sufficiently large. Furthermore, let us assume that we know the number of sources {$ L > 1 $} a priori. The DoA estimation task can then be expressed as a multilabel-multiclass classification problem, also referred to as multidimensional classification, cf., e.g, \cite{Zhang2014,Herrera2016}, i.e., {for each input vector $ \bm{x} $ we try to find $L$ correct labels $g(1),\dots,g(L) $ out of $ G $ sectors (classes).}
	
	A direct extension of the multiclass formulation from the single source case to multiple sources leads to the label power-set method \cite{Boutell2004}. There, we assign each combination of sectors to one class. Since we do not care for the order of the DoA estimates and two sources cannot lie in the same sector, for modeling the DoA estimation problem with the label power-set method it suffices to consider all unordered $L$-tuples of sectors without repetition, i.e., the necessary number of classes is given by the binomial coefficient $G\choose L$. This means that, for $G\gg L$, the number of classes, which is equal to the number of neurons in the output layer, grows exponentially with the number of sources. Hence, the power-set method quickly becomes computationally infeasible.
	
	A more tractable alternative is the binary relevance (BR) approach \cite{Boutell2004}, which has been used for DoA estimation in \cite{Chakrabarty2019,Yao2020, Papageorgiou2020a}. The idea is to train a network that realizes $G$ binary classifiers, i.e., one for each sector, where the $g$-th classifier tries to produce the probability that a given data sample stems from a scenario where there is an active source in the $g$-th sector. {To this end, instead of $L$-tupels of labels, we introduce the label vectors $\bm{v} = [v_1,\dots,v_G] \in \{0,1\}^G $ of length $G$, which have a one at each element that corresponds to an active sector and zeros elsewhere.}
	 To obtain outputs {$z_g,g=1,\dots,G,$} between $0$ and $1$ that can be interpreted as probabilities, a sigmoid activation is employed at every output neuron. The loss function for the whole network is the sum of the binary cross-entropies of each binary classifier, i.e.,
	\begin{equation}
	{
	-\sum\limits_{g=1}^G v_g\log\left(z_g\right)+(1-v_g)\log\left(1-z_g\right).}
	\end{equation}
	
\begin{figure}[h]
	\begin{center}
		\includegraphics{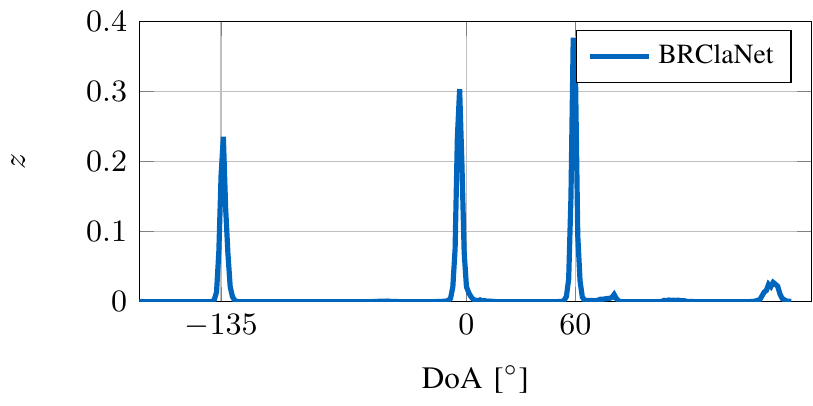}
	\end{center}
	\vspace{-5pt}
	
	\caption{Exemplary Classifier Output for Three Equally Powered Sources at $\bth=[-135^\circ, 0^\circ, 60^\circ]$ at $20\,\text{dB}$ SNR and $N=10$.}
	\label{fig:classoutput}
\end{figure}

	{At the core of the BR method lies the assumption of stochastically independent labels. Hence, the output of the $g$-th binary classifier is formed independently from the decision of the other $G-1$ binary classifiers. By this, the plain BR approach can not properly model correlations between the labels. This does not only concern prior knowledge on the distribution of the DoAs but also the number of sources $L$ is not directly taken into account by the design of the NN architecture.\footnote{In scenarios where the model order is not known a priori, this can be an advantage, as a single NN can be trained for multiple model orders and the model order selection can be accomplished by the same NN, e.g., using thresholding of the output.} The $g$-th classifier does not consider the output of the classifier for sector $g',g'\neq g$, which might indicate that a source lying in sector $g'$ is much more probable than in sector $g$. For the considered DoA estimation problem, the outputs between two binary classifiers corresponding to neighboring sectors are highly correlated. This can be seen in the Figure~\ref{fig:classoutput}, which shows the outputs $\bm{z}$ of the trained NN for an exemplary realization of three sources based on the BR method for $ G = 288 $. Therefore, in practice a consecutive peak search is required to obtain the DoA estimates from the spectrum formed by the classifier outputs.}
	
	Another approach to multilabel classification that is able to cope with dependencies between labels are binary classifier chains \cite{Read2011,Read2020}. A common embodiment of one of these binary classifier chains consists of a concatenation of $ G $ binary classifiers, one for each {sector $ g = 1,\dots,G $}, where each node in this chain bases its decision on the original input data sample and the output of the previous stages. A typical problem with such an architecture is that its performance is heavily dependent on the order of the binary classifiers. Therefore, the averaging of the outputs over an ensemble of binary classifier chains with different orders have been investigated in \cite{Read2011}. 
	
	\section{ChainNet}
	\label{sec:chainnet}
\begin{figure*}
	\begin{center}
		\includegraphics{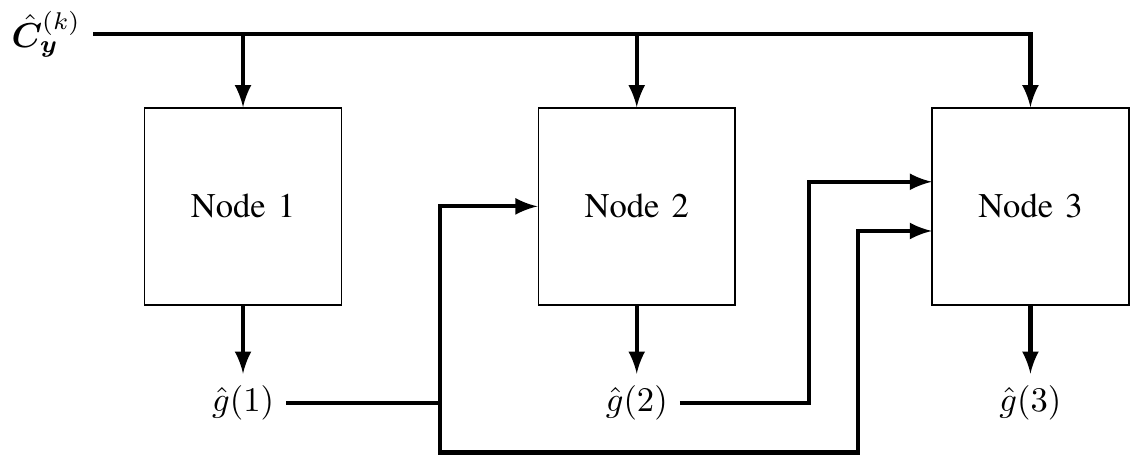}
	\end{center}
	\caption{Illustration of ChainNet Architecture for $L=3$.}
	\label{fig:chainnet}
\end{figure*}

	For the DoA estimation problem, we propose a new architecture, which has been heavily inspired by the binary classifier chains. As we have a multilabel-multiclass classification problem, where we know the number of true labels $L$ a priori, we can construct a classifier chain consisting of $L$ consecutive multiclass classifiers. This means that each stage of this multiclass classifier chain shall provide an estimate for one of the $L$ sources. In other words, the proposed classifier chain estimates the DoAs {successively}. To this end, each multiclass classifier in the chain has---apart from the input data---the structure of a multiclass classifier for a single source as discussed above, i.e., a neural network with $G$ output neurons, whose outputs are subject to a softmax activation and is trained based on a cross-entropy loss function (\ref{eq:crossentropy}). 
	{Unlike common binary classifier chains, the proposed approach concatenates a smaller number of single source multiclass classifiers equal to the number of labels $ g(1),\dots,g(L) $ to be estimated.}
	Due to simplicity and to maintain a high comparability to the recently proposed NN-based DoA estimators MCENet \cite{Barthelme2020b} and GramNet \cite{Barthelme2021}, we consider a fully connected, feedforward architecture consisting of $N_{\text{h}}$ hidden layers with $N_{\text{u}}$ neurons {with a rectified linear unit (ReLU) activation function} per layer for each stage. The $\ell$-th stage in the chain is fed with the original sample covariance matrices $\hat{\bm{C}}_{\bm{y}}^\supk$ of all subarrays $ k = 1,\dots,K $ and the discrete DoA estimates of the previous nodes {$ \hat{g}(1),\dots, \hat{g}(\ell-1) \in \{1,\dots,G\} $}, cf. Figure~\ref{fig:chainnet}. We refer to such a multiclass classifier chain as \emph{ChainNet}.
	
	Each single multiclass classifier shall produce the DoA estimate for one of the sources, however, it is still unclear which source should be estimated by which classifier. In our case, therefore we propose to estimate the sources in a predefined order of their true DoAs, {i.e., $ g(1) < g(2) < \dots < g(L) $ according to a corresponding numbering of the sources.}
	For the training of the individual nodes, we can follow two different paths. 
	
	\emph{i)} One way is training the first classifier as for the single source case. Then the first node of the classifier chain is used to estimate the first DoA {for each element} of the whole training set. The resulting estimates {$ \hat{g}_n(1) $} along with the subarray sample covariance information is subsequently used in the second node, and so on and so forth. 
	
	\emph{ii)} Another option is to train all nodes simultaneously with the assumption that the estimates from the previous nodes are perfect. In that case, the training data for the second node again consists of the sample covariance matrices of the subarrays and the true information of which sector comprises the first source, in the form of the corresponding {known sector labels $g_i(1)$}. This variant does not only allow to train all the nodes simultaneously, but provided a better estimation accuracy in our simulations.
	
	For the design of the training set, we use data from the underlying stochastic model (\ref{eq:smldist}). Thereby, the model order $L$, which we assume to be known a priori, is fixed. To ensure that two sources do not lie in the same sector, for each realization, we first select $L$ distinct sectors randomly and draw the DoA within each sector from a uniform distribution. According to (\ref{eq:smldist}), the noise and transmit signal realizations follow a complex normal distribution. Furthermore, we apply some kind of data preprocessing by fixing the power of the strongest source to $\sigma_{s,\text{max}}^2=1$, which works as a data normalization. The power of each weaker source in decibel is drawn from a uniform distribution between $0\,\text{dB}$ and $\sigma_{s,\text{min}}^2$. This ensures that the power of a source cannot become arbitrarily small, which would effectively reduce the model order. The noise power is uniformly distributed between $\sigma_{\eta,\text{min}}^2$ and $\sigma_{\eta,\text{max}}^2$ as well. With these settings, we can produce arbitrarily many data samples, each consisting of $KN$ i.i.d.\ received signal realizations $\bm{y}^{(k)}(n),n=1,\dots,N, k=1,\dots,K$, for the training set. This allows us to feed new, previously unseen realizations to the NN in each step of the gradient descent of the learning algorithm, which makes the training inherently robust towards overfitting. As we know the true DoAs for each data sample, the proposed NN estimators are based on supervised learning.
	
	As input data, we do not directly pass the complex-valued received signal realizations to the NN, since state-of-the-art machine learning frameworks such as TensorFlow \cite{Tensorflow2015} can only model real-valued NNs. One way to overcome this problem is passing the real and imaginary parts of the received signals to the NN. However, we use a different kind of preprocessing of the input data. As has been shown in \cite{Costa1999,Niu2017,Barthelme2020}, sample covariance matrix information is a suitable format for the input data of NNs when we are working in the DoA context. This is not very surprising, as the stochastic model of the input data (\ref{eq:smldist}) is uniquely parameterized by the subarray covariance matrices. Hence, we first form the $K$ subarray sample covariance matrices $\hat{\bm{C}}_{\bm{y}}^{(k)}$, $k=1,\dots,K$, from the received signal realizations. Then, we stack their real parameters, i.e., their diagonal elements and the real and imaginary parts of their upper triangle, in one large vector per data sample.
	
	\section{Comparison with MAP Estimator}
	As discussed in Section~\ref{sec:singlesource}, a NN trained on the cross-entropy objective in combination with a softmax activation layer for the single source case constitutes an approximation of the MAP estimator. With the same argument, the computationally infeasible label power-set method leads to a MAP approximation for the multiple source case. In contrast, a MAP interpretation of classifiers based on the binary relevance method does not seem obvious. Thus, in this section, we now compare the ChainNet approach with the theoretical optimal MAP estimator by deriving a reformulation of the MAP estimator as a successive estimation task. 
	
	The MAP estimator maximizes the posterior probability mass function
	\begin{equation}
	p({\theta}_1,\dots,{\theta}_L|\bm{y}),
	\end{equation}
	where, under the condition that the observations $\bm{y}$ are given, the discrete value of ${\theta}_\ell \in \{1,\dots,G\} $ indicates the potential sector in which the $\ell$-th source is located.\footnote{{Please note that abusing the notation, but for the generality of the argument, the angular variable $ \theta_\ell $ is used to denote the sector label $g(\ell)$.}} The MAP estimate is therefore given by
	\begin{align}
	\hat{{\theta}}^{\text{MAP}}_{1},\dots,\hat{{\theta}}^{\text{MAP}}_{L}&=\argmax_{{\theta}_1,\dots,{\theta}_L}p({\theta}_1,\dots,{\theta}_L|\bm{y})\label{eq:MAP}\\
	&=\argmax_{{\theta}_1,\dots,{\theta}_L}p(\bm{y}|{\theta}_1,\dots,{\theta}_L) p({\theta}_1,\dots,{\theta}_L)\nonumber\\
	&=\argmax_{{\theta}_1,\dots,{\theta}_L}p(\bm{y}|{\theta}_1,\dots,{\theta}_L) \prod_{\ell=1}^{L}p({\theta}_\ell|{{\bth}}_{\ell-1}),\nonumber
	\end{align}
	where {${\bth}_{\ell-1} = [{\theta}_{1},\dots,{\theta}_{\ell-1}]$.}
	
	{
	This MAP estimator can be reformulated as a successive estimation task. To that end, let us assume w.l.o.g.\ that the DoAs are in ascending order. We can rewrite the optimization problem for the first source as
	\begin{align}
	\hat{\theta}^{\text{MAP}}_{1}
	&=\argmax_{{\theta}_1}\,f_1({\theta}_1;\bm{y}),
	\label{eq:opt1} 
	\end{align}
	with 
	\begin{align}
	f_1({\theta}_1;\bm{y})&=\max_{{\substack{{\theta}_L,\theta_{L-1},\dots,{\theta}_2\\
			{\theta}_L>,\theta_{L-1}>\dots>{\theta}_2>{\theta}_1}}} p({\theta}_2,\dots,{\theta}_L|{\theta}_1;\bm{y}) p(\theta_1). \label{eq:f1}
	\end{align}
	Furthermore, under the assumption that we know the MAP estimates for first $\ell-1$ sources, we obtain the MAP estimate of the $\ell$-th source by
	\begin{align}
	\hat{{\theta}}^{\text{MAP}}_{\ell}&=\argmax_{{\theta}_\ell>\hat{{\theta}}^{\text{MAP}}_{\ell-1}}\,f_\ell({\theta}_\ell;\bm{y},\hat{{\theta}}^{\text{MAP}}_{1},\dots,\hat{{\theta}}^{\text{MAP}}_{\ell-1}), \label{eq:gl}
	\end{align}
	with $f_\ell({\theta}_\ell;\bm{y},\hat{{\theta}}^{\text{MAP}}_{1},\dots,\hat{{\theta}}^{\text{MAP}}_{\ell-1})$ given by
		\begin{multline}
	 \max_{{\substack{{\theta}_{L},\theta_{L-1},\dots,{\theta}_{\ell+1} \\
			{\theta}_L>\theta_{L-1}>\dots>{\theta}_{\ell+1}>{\theta}_\ell}}} p({\theta}_{\ell+1},\dots,{\theta}_L|{\theta}_{\ell};\bm{y},\hat{{\theta}}^{\text{MAP}}_{1},\dots,\hat{{\theta}}^{\text{MAP}}_{\ell-1})
			\\ \cdot p({\theta}_{\ell};\bm{y},\hat{{\theta}}^{\text{MAP}}_{1},\dots,\hat{{\theta}}^{\text{MAP}}_{\ell-1}).
	\end{multline}
	}
	
	Note that finding the MAP estimate of the first source in (\ref{eq:opt1}) would inherently yield all the MAP estimates of the other sources, since for each ${\theta}_1$, we search for the ${\theta}_2,\dots,{\theta}_L$ in (\ref{eq:f1}) that maximize the posterior probability. Therefore, the optimization problem is nothing else than a reformulation of the original MAP estimation problem in (\ref{eq:MAP}). {But because of this reformulation,  and knowing the true DoAs of all sources of the available training set,\footnote{{Refer to the 2$^\text{nd}$ option in Section \ref{sec:chainnet}.}} we are able to directly identify the learning task for estimating the first source. Consequently, there is a function to be learned} that maps the observations $\bm{y}$ to the MAP estimate of the first source, namely the function 
	\begin{equation}\argmax_{{\theta}_1}\,f_1({\theta}_1;\bm{y}).
	\end{equation}
	Similarly, we see that the MAP estimate for the $\ell$-th source can be written as a function {to be learned} of the observations $\bm{y}$ and the previous estimates, i.e., 
	\begin{align}
	\displaystyle\argmax_{{\theta}_\ell>\hat{{\theta}}^{\text{MAP}}_{\ell-1}}\,f_\ell({\theta}_\ell;\bm{y},\hat{{\theta}}^{\text{MAP}}_{1},\dots,\hat{{\theta}}^{\text{MAP}}_{\ell-1}).
	\label{eq:UnivApprox}
	\end{align}

	The following argumentation is based on the universal approximation theorem for a sufficiently large neural networks \cite{Cybenko1989,Hornik1991}. Based on this theorem, each stage of the ChainNet is designed to approximate the corresponding function in (\ref{eq:UnivApprox}). To this end, the respective cross-entropy loss function of the $\ell$-th stage and its training data must be selected accordingly. Consequently, the gradient descent attempts to increase the value of exactly the output corresponding to the sector containing the $\ell$-th source, while simultaneously decreasing the outputs of all other sectors based on observations $\bm{y}$ and the known parameters $ {\theta}_1,\dots,{\theta}_{\ell-1} $ of the sources located in the predefined order before.
	In other words, the target is to obtain a correct detection of the sector that contains the $\ell$-th source. The proposed method approximates the MAP estimator, which in general maximizes the probability of a correct detection \cite{Stoica2004}.
	
	{Finally, we would like to briefly discuss the role of the introduced estimation order. While the order of the DoA angles to be estimated does not necessarily play a role in the reasoning presented in this section, it has been found that an order based on the size of the angles is, however, very beneficial for the supervised learning of the individual nodes of the classifier chain, since the estimates of the previous nodes provide strong prior information about the range of labels to be estimated by the subsequent nodes.} {In contrast, we also explored an ordering of sources by their aggregated instantaneous transmit power. However, this approach resulted in very poor estimation accuracy for more than two sources, which is an indication that successive DoA estimation by the ChainNet approach is not related to the well-known class of generalized expectation maximization algorithms, cf. for example the SAGE (space-alternating generalized expectation-maximization) algorithm \cite{Fleury99}.} 
	
	\section{Simulation Results}
	In our simulations, we compare the achievable accuracy of the proposed ChainNet approach with that of the sparse recovery technique SPICE \cite{Stoica2011,Suleiman2018}, the BRClaNet method \cite{Chakrabarty2019}, and the recently presented MCENet regression network \cite{Barthelme2020b}. Additionally, we add the results for the MUSIC estimator \cite{Schmidt1986} in the case of fully sampled arrays, and the GramNet estimator \cite{Barthelme2021} for arrays with subarray sampling. In all cases, we show the performance of the plain estimators with {discrete estimates} and their respective off-grid extensions, i.e., a combination of an initialization with the plain estimator output and subsequent gradient steps on the likelihood function \cite{Barthelme2020b}.
	
	We consider a uniform circulcar array (UCA) with an azimuthal field of view from $0$ to $2\pi$. For a UCA with $M$ omnidirectional antennas, the array manifold is given by
	\begin{equation}
	\bm{a}_\text{UCA}(\theta)=\left[\begin{array}{l}
	\alpha_\text{UCA}(\phi)^{\cos(\theta)}\\
	\alpha_\text{UCA}(\phi)^{\cos(\theta-\frac{2\pi}{M})}\\
	\quad\quad\vdots\\
	\alpha_\text{UCA}(\phi)^{\cos(\theta-\frac{2\pi(M-1)}{M})}\\
	\end{array}\right],
	\label{eq:ucasteering}
	\end{equation}
	with
	\begin{equation}
	\alpha_\text{UCA}(\phi) = \exp\left(-\text{j}2\pi \frac{R}{\lambda}\cos(\phi)\right).
	\end{equation}
	where $\theta$ is the azimuth angle and $\phi$ is the elevation angle to the source, $R$ denotes the array radius, and $\lambda$ is the wavelength of the impinging electro-magnetic wave. In the following, we assume that all the sources lie in the same plane as the antenna elements, such that the elevation $\phi$ is $0$ for all sources and the DoAs are fully described by the azimuth angles $\bth = [\theta_1,\dots,\theta_L] $. The considered array consists of $M=9$ antennas and the ratio of radius to wavelength is $1$. 
	
	In general, we use an oversampling factor $Q=32$ for the DoA estimation algorithms. This means that for SPICE the field of view is sampled by $Q$ times $M$ is $G=288$ grid points. Similarly, the MUSIC spectrum is evaluated on the same grid and the number of sectors for the classification-based ChainNet and BRClaNet approaches is also $Q\cdot M$. 
	
	For each stage of the ChainNet-NN and for the MCENet and BRClaNet NNs, the architecture is identical apart from the input and output layers. The common parameters for the hidden layers, the training set, and the optimizers is summarized in Table~\ref{tab:NNparam}. The respective dimensions of the input and output layers of each network can be easily inferred from the parameter values provided above.
	
	\begin{table}
		\caption{Simulation Parameters DoA Estimation}\label{tab:NNparam}
		\begin{center}
			\begin{tabular}{c|c}
				Parameter & Value\\\hline
				$\sigma_{s,\text{min}}^2$ & $-9\,\text{dB}$\\
				$\sigma_{\eta,\text{min}}^2$ & $-10\,\text{dB}$\\
				$\sigma_{\eta,\text{max}}^2$ & $30\,\text{dB}$\\
				$N_{\text{h}}$ & $4$\\
				$N_{\text{u}}$ & $4096$\\
				Weight Initialization & Glorot \cite{Glorot2010}\\
				Batch Size & $256$\\
				Optimizer & Adam \cite{Kingma2014}\\
				Learning Rate & $10^{-4}$\\
				Samples per Training Set & $64\cdot 10^6$\\
			\end{tabular}
		\end{center}
	\end{table}
	
	For our test sets, we use realizations from $L$ equally strong sources. Here, the DoAs of each realization are drawn from a uniform distribution over the entire field of view, the same distribution from which the training data were drawn. This allows us to make a fair comparison between the NN-based and model-based estimators, since the performance of the NN-based methods for a single DoA is affected by its representation in the training set. However, with changing DoAs for each realization and a limited number of Monte Carlo runs, the Cram\'er-Rao bound (CRB) is no longer a useful limit for the achievable estimation error due to its stochastic nature (cf. \cite{Barthelme2020b}). Instead, we consider a so-called Genie ML scheme, which is based on the standard ML approach, but where the solver based on a local gradient descent is initialized with the true DoAs. 
	
	In the following, we will first investigate a scenario with subarray sampling, which matches the simulation scenario of \cite{Barthelme2020b} and \cite{Barthelme2021}. Here, we study the case of uncorrelated and correlated sources. Then, we extend our analysis to fully sampled arrays.
	
	\subsection{Subarray Sampling - Uncorrelated Sources}
\begin{figure}[t]
\begin{subfigure}{\columnwidth}
	\centering
	\includegraphics{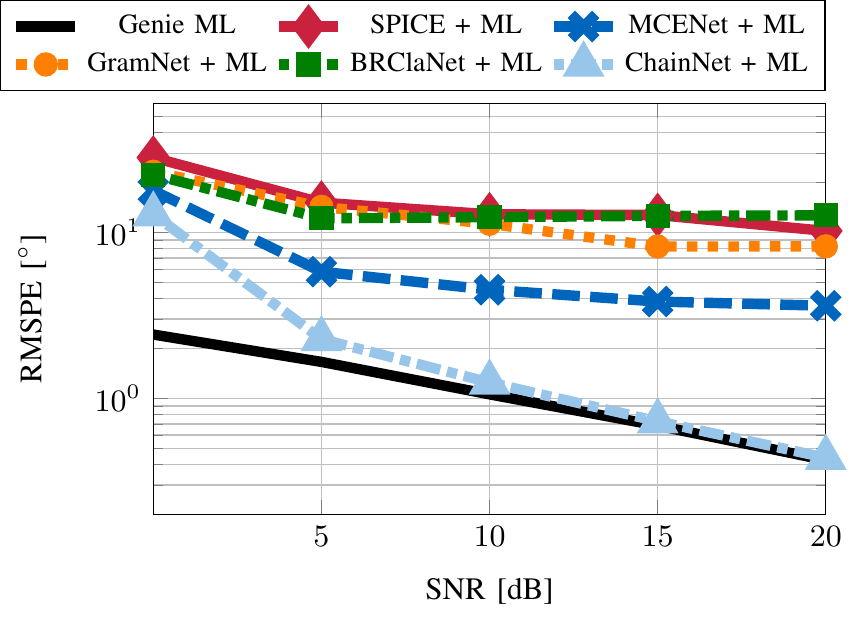}
	\vspace{-5pt}
	
	\caption{RMSPE vs.~SNR}
	\label{fig:rmspeL2}
\end{subfigure}

\begin{subfigure}{\columnwidth}
	\centering
		\includegraphics{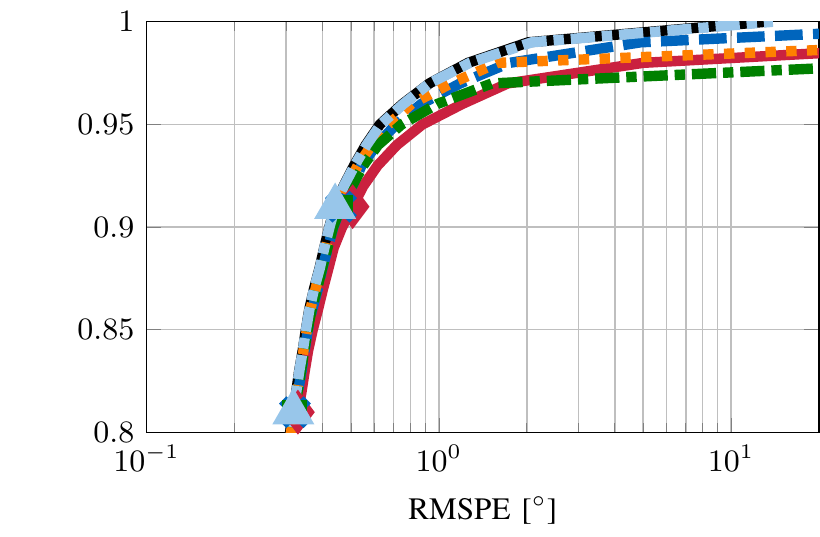}
	\vspace{-5pt}
	
	\caption{Empirical Cumulative Density Function at SNR$=20\,\text{dB}$.}
	\label{fig:cdfL2}
\end{subfigure}
\caption{Simulation Results for $L=2$, $M=9$, $ K = 4 $, $ W=3 $, $N=10$, with gradient-based refinement.}
\end{figure}

	Let us consider a scenario with $W=3$ RF chains, which are used to sample $K=4$ subarray configurations. The selected antennas per subarray are given in Table~\ref{tab:subsampling}, which uses a clockwise numbering of the antenna elements of the UCA. The achievable root mean square periodic error (RMSPE) given by
	\begin{equation}
	\text{RMSPE}=\sqrt{\frac{1}{L}\sum_{\ell=1}^L \Exp_\theta\left[\left|\modulo_{[-\pi,\pi)}\left(\theta_\ell-\hat{\theta}_\ell\right)\right|^2\right]}.
	\end{equation}
	of the different estimators have been obtained by means of Monte Carlo simulations with $10^4$ realizations, respectively. For $L=2$, the results for different SNR are shown in Figure~\ref{fig:rmspeL2}. We observe that the proposed ChainNet approach outperforms all other estimators and is the only estimator that is able to achieve the Genie ML lower bound when a gradient-based refinement step is applied at the end. All other estimators have fairly drastic DoA estimation errors for some realizations, and these outliers can have a significant impact on the RMSPE results presented. In this sense, we can deduce that the proposed ChainNet approach provides the most robust initializations for the successive gradient steps of all the estimators studied. {The ChainNet approach even leads to fewer outliers than the MCENet estimator, whose loss function specifically promotes robustness against outliers \cite{Barthelme2020b}.} BRClaNet, the other classification-based estimator, does not come close to the proposed ChainNet approach, as it only achieves an overall performance similar to the SPICE and GramNet estimators. Note that both BRClaNet and GramNet utilize a peak search to obtain the DoA estimates from an estimated spectrum. They suffer especially from closely spaced sources. However, even considering a minimum separation between two sources does not eliminate all outlier realizations for the BRClaNet approach in contrast to the GramNet method (cf.~\cite{Barthelme2021}). 
	
	\begin{table}[h]
		\caption{Subarray Sampling Scheme}\label{tab:subsampling}
		\begin{center}
			\begin{tabular}{c|c}
				$k$ & Antenna Elements\\\hline
				$1$ & $1,\;2,\;9$\\
				$2$ & $1,\;3,\;8$\\
				$3$ & $1,\;4,\;7$\\
				$4$ & $1,\;5,\;6$
			\end{tabular}
		\end{center}
	\end{table}
	
\begin{figure}[t]
\centering
\begin{subfigure}[t]{\columnwidth}
	\centering
	\includegraphics{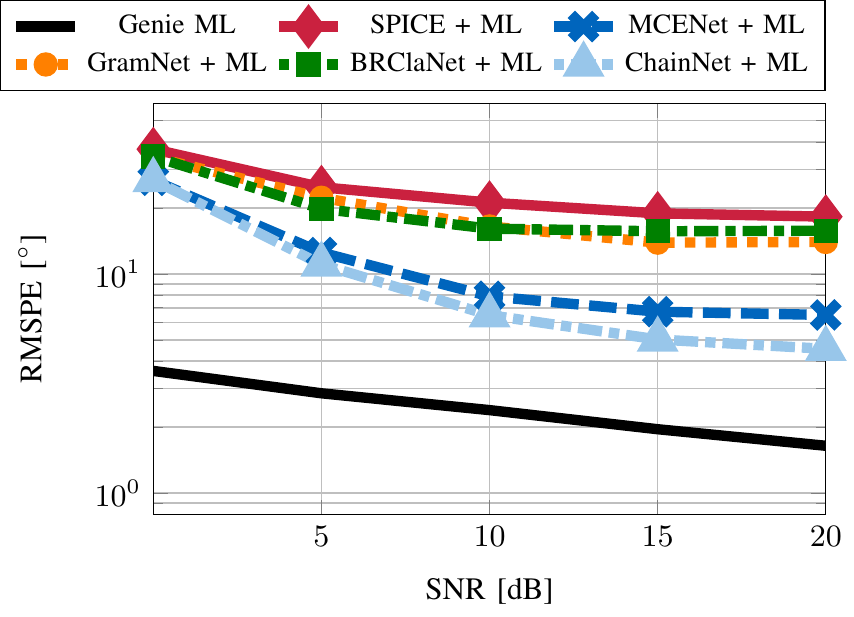}
	\vspace{-5pt}
	
		\caption{RMSPE vs.~SNR}
	\label{fig:rmspeL3}
\end{subfigure}

\begin{subfigure}[b]{\columnwidth}
\centering
		\includegraphics{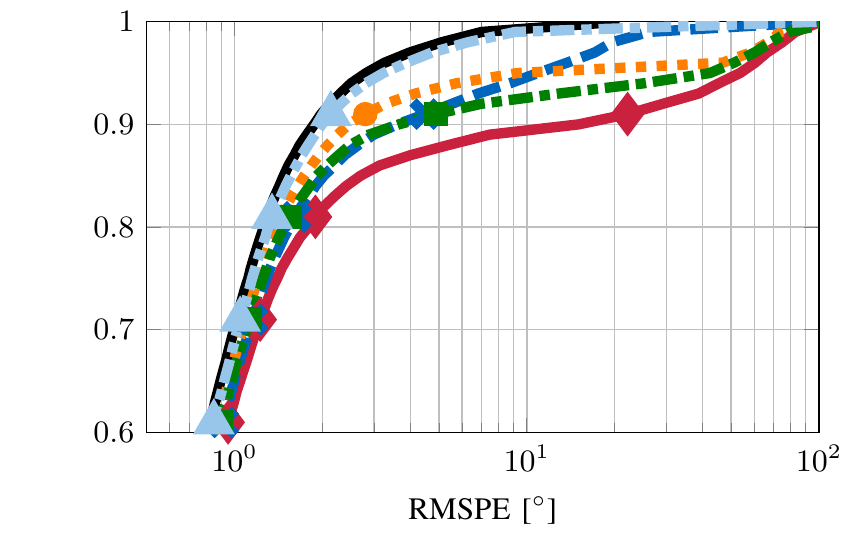}
	\vspace{-5pt}
	\caption{Empirical CDF at SNR$=20\,\text{dB}$}
	\label{fig:cdfL3}
\end{subfigure}	
\caption{Simulation Results $L=3$, $M=9$, $ K = 4 $, $ W=3 $, $N=10$, with gradient-based refinement.}
\end{figure}

	For $L=3$, Figure~\ref{fig:rmspeL3} shows a similar outcome. Again, we see that the ChainNet approach yields the most robust estimates. However, in this harder scenario, where the number of sources is equal to the number of RF chains, also ChainNet suffers from outliers. To be able to assess the robustness of the different methods in more detail, we show a cutout of the empirical cumulative density function (CDF) of the hybrid estimators in Figure~\ref{fig:cdfL3}. From the CDF, we can see that the proposed novel estimator achieves the Genie ML performance in about $99\%$ of realizations.
	
\begin{figure}[h]
	\begin{center}
		\includegraphics{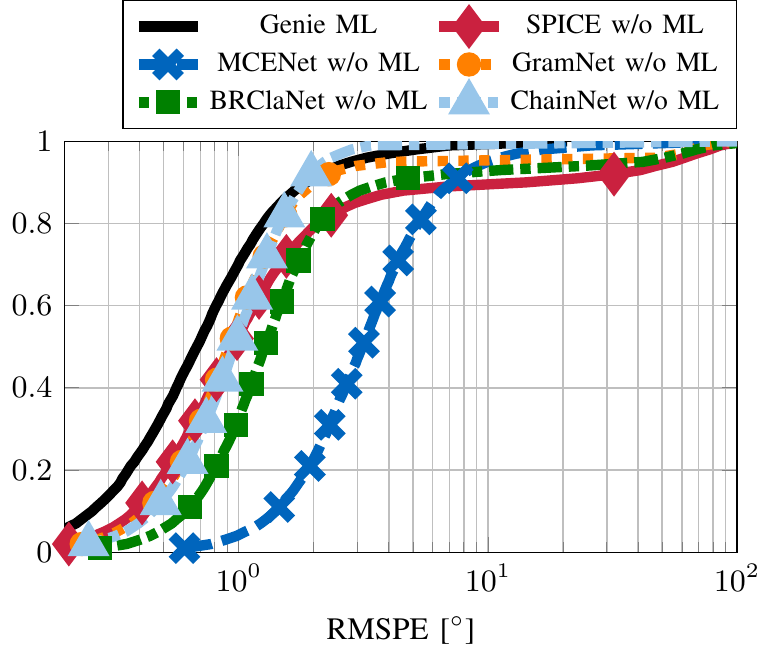}
	\end{center}
	\vspace{-5pt}
	
	\caption{Empirical CDF, SNR$=20\,\text{dB}$, $L=3$, $M=9$, $ K = 4 $, $ W=3 $, $N=10$, {\emph{without} gradient-based refinement.}}
	\label{fig:cdfL3grid}
\end{figure}

	Its main advantage over the second-best MCENet-NN can best be examined by looking at the performance of the ChainNet approach without any refinement, i.e., without consecutive gradient steps, cf.\ the CDF plot in Figure~\ref{fig:cdfL3grid}. The MCENet approach is not able to achieve a very high accuracy for the non-outlier realizations by itself due to its learning objective that trades off a high robustness against accuracy for these cases \cite{Barthelme2020b}. In contrast, the ChainNet approach does not only achieve a superior robustness, but also produces very accurate estimates that can keep up with the SPICE and GramNet estimator for the non-outlier realizations.
	
\begin{figure}[h]
	\begin{center}
		\includegraphics{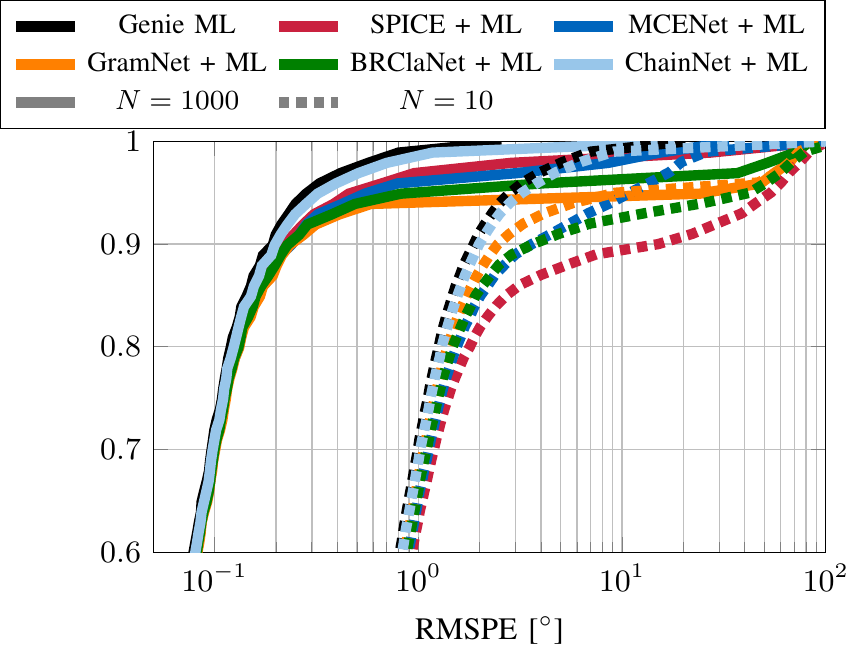}
	\end{center}
	\vspace{-5pt}
	
	\caption{Comparison of Empirical Cumulative Density Functions for Varying $N$ at SNR$=20\,\text{dB}$ for $L=3$, $M=9$, $K = 4 $, $ W=3 $, with gradient-based refinement.}
	\label{fig:cdfL3N=10vs1000}
\end{figure}

	What we have not discussed yet, is how the different algorithms behave for a varying number of snapshots. To that end, we first take a look at Figure~\ref{fig:cdfL3N=10vs1000}, which shows the empirical CDFs for $N=1000$ snapshots as solid lines. The CDF plot for $N=10$, shown in Figure~\ref{fig:cdfL3}, has been added to Figure~\ref{fig:cdfL3N=10vs1000} as dashed lines. From the CDFs for $N=1000$, we can see that more snapshots improve not only the overall achievable RMSPE (shift to the left), but we have fewer outliers than for $N=10$. Especially, the SPICE method profits heavily from the increased number of observations. This behavior of the SPICE estimator is not surprising, as it is based on a covariance-matching criterion, similar to the GLS estimator, which has been proven to be a consistent estimator (for a sufficiently dense grid) \cite{Sheinvald1999}. For a high number of snapshots $N$, the sample covariance matrices are consistent estimates of the true subarray covariance matrices, which again justifies the validity of the covariance-matching objective.
	
\begin{figure}[h]
	\begin{center}
		\includegraphics{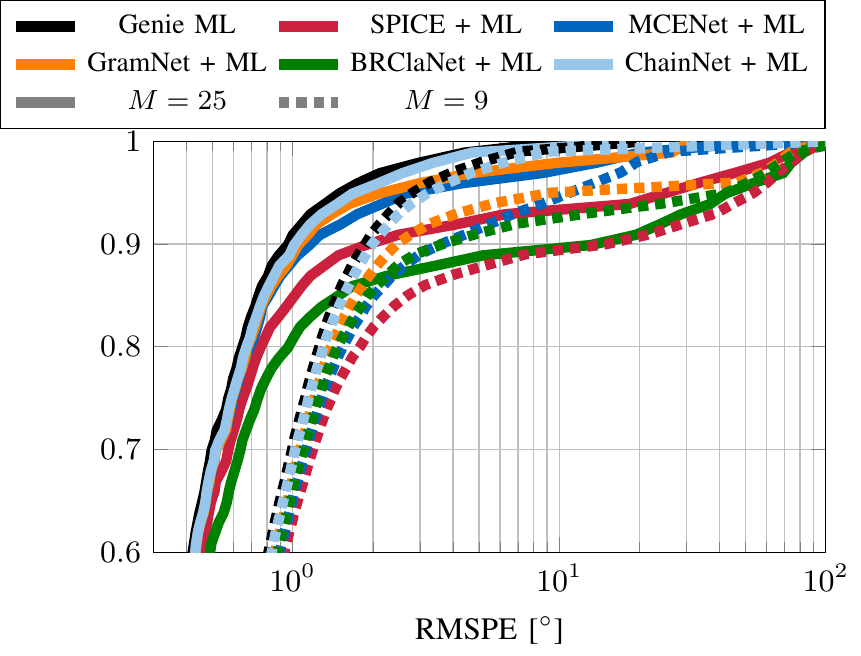}
	\end{center}
	\vspace{-5pt}
	
	\caption{Comparison of Empirical Cumulative Density Functions for Varying $M$ at SNR$=20\,\text{dB}$ for $L=3$, $ K = 4 $, $ W=3 $, $N=10$, with gradient-based refinement.}
	\label{fig:cdfL3M=9vs25}
\end{figure}

	In a similar fashion to Figure~\ref{fig:cdfL3N=10vs1000}, the CDFs for a differing number of antenna elements are compared in Figure~\ref{fig:cdfL3M=9vs25}. The plot shows the results for an antenna array with $M=25$ antennas and $W=3$ RF chains in comparison to the previously considered $9$-element antenna array. Again, we observe an improvement of the overall RMSPE and a decline in the number of outliers. However, $25$ antennas are not enough to fully close the gap between the SPICE algorithm and the NN-based methods. Note that with the number of antennas, the number of subarrays that need to be sampled increases as well. Depending on the time that is required to switch between two subarray constellations, a trade-off between more antenna elements and an increased number of snapshots is necessary.
	
	\subsection{Subarray Sampling - Correlated Sources}
	Another interesting scenario arises for correlated sources. There, the positive semidefinite covariance matrix of the transmit signals $\bm{C}_{\bm{s}}$ is no longer diagonal. Instead, we consider that the covariance matrices for equal transmit powers are parameterized as follows 
	\begin{equation}
	\bm{C}_{\rho} = \begin{bmatrix}
	1 & \rho & \rho^2\\
	\rho & 1 & \rho\\
	\rho^2 & \rho & 1,
	\end{bmatrix}
	\end{equation}
	with the correlation coefficient $\rho\in[0,1]$. Adding differences in the transmit powers of the individual sources $\bm{\lambda}$, this model covers all covariance matrices of the form
	\begin{equation}
	\bm{C}_{\bm{s}}=\bm{C}_\rho^{1/2}\diag(\bm{\lambda})\bm{C}_\rho^{1/2}.
	\end{equation}
	Although this model does not comprise all possible cases of correlated sources, the correlation factor $\rho$ provides an intuitive measure of the degree of correlation. Furthermore, for the creation of the training set, it allows us to sample from the covariance matrices in an easy way by drawing $\rho$ according to a uniform distribution in addition to the transmit powers $\bm{\lambda}$, cf.\ Section~\ref{sec:chainnet}.
	
	To maintain the excellent performance of the presented NNs for uncorrelated sources, we chose to use NNs trained for uncorrelated sources as an initialization. Then, we use 16 million training data samples from the correlated data model to adapt the existing networks to the new scenario.\footnote{{In general, nothing prevents us from training the NN from scratch with data from the correlated system model. However, we observed that for a uniformly drawn $\rho$, the resulting network's performance for uncorrelated sources is no longer significantly better than for other estimators. By using the uncorrelated case as a baseline, we retain the superior performance in the uncorrelated case, while the performance degradation of ChainNet for correlated sources is negligible compared to training directly on correlated data.}} This procedure is similar to the online training method presented in \cite{Barthelme2020} for model order selection, which showed that training a NN on an artificial data model and adapting the NN to unseen data by only a few online training steps can be feasible. Moreover, this approach has been studied in \cite{Barthelme2020b} for the MCENet estimator, where it also showed very good results. Note that the SPICE estimator is inherently based on the uncorrelated data model, and therefore, suffers from a model mismatch in the correlated source scenario, which cannot be overcome by simply augmenting a training data set. For comparison, we also include the results of a ChainNet-NN, which has been trained on data with uncorrelated sources only. This allows us to assess how well the ChainNet approach can generalize to previously unseen data without an augmentation of the training set.
	
\begin{figure}[h]
\begin{subfigure}{\columnwidth}
	\centering
		\includegraphics{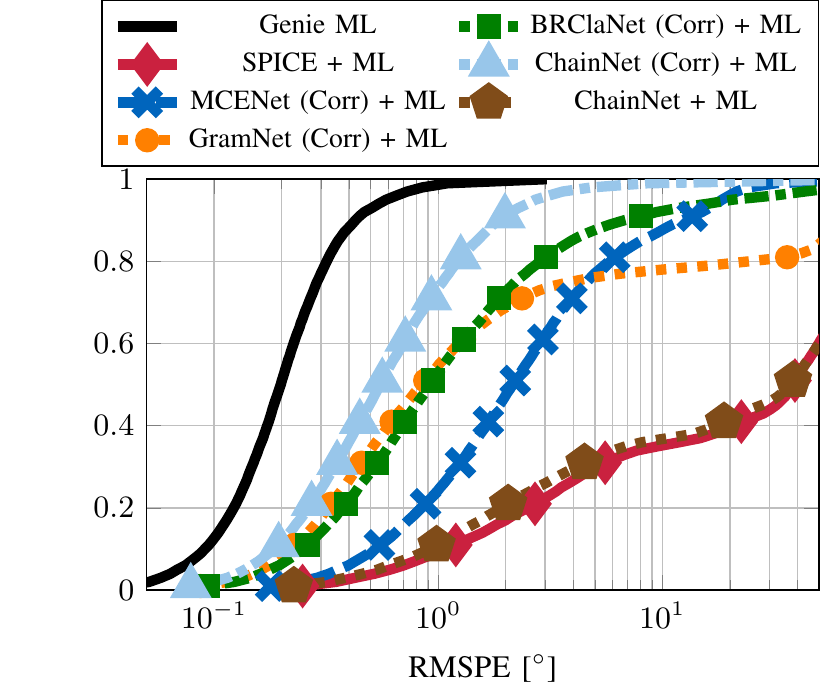}
	\vspace{-5pt}
	\caption{Empirical CDF for Correlated Sources with $\rho=1$ at SNR$=20\,\text{dB}$.}
	\label{fig:cdfCorr}
\end{subfigure}

\begin{subfigure}{\columnwidth}
	\centering
		\includegraphics{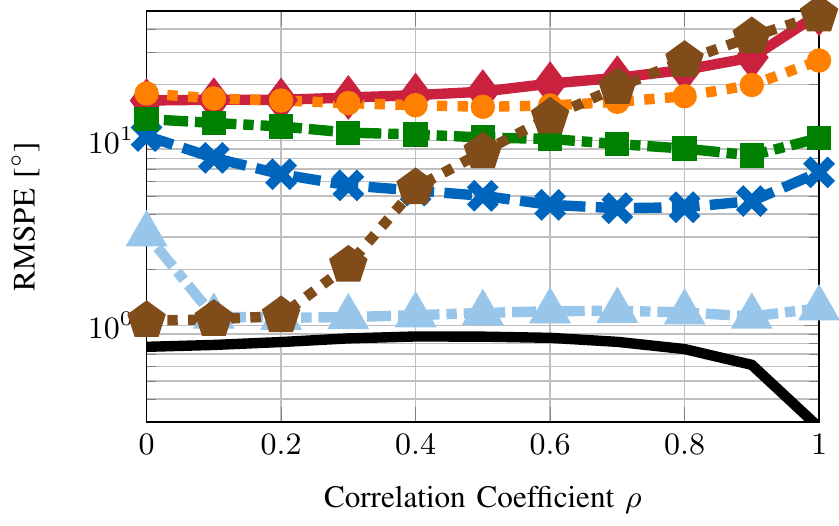}
	\vspace{-5pt}
	
	\caption{RMSPE vs.~Correlation Coefficient $\rho$ for the Top $99\%$ of Realizations.}
	\label{fig:corr}
\end{subfigure}
\caption{Simulation Results for Correlated Sources for $L=3$, $M=9$, $ K = 4 $, $ W=3 $, $N=10$, with gradient-based refinement.}
\end{figure}

	Figure~\ref{fig:cdfCorr} shows the results for fully correlated sources, i.e. $\rho = 1$, at an SNR of $20$\,dB. We see that the ChainNet approach, which has been adapted to the correlated source case, is again superior to all other estimators and comes closest to the Genie ML bound. On the contrary, the SPICE reference and the ChainNet-NN that has been trained for uncorrelated sources only are not able to provide reasonable estimates in this case. In Figure~\ref{fig:corr}, we show the performance of the different estimators for varying $\rho$. Hereby, we consider only the $99\%$ of realizations with the smallest errors for the presented RMSPE to discard some of the outlier realizations. The results show that the ChainNet-NN, which is based on the appropriate training set, is able to achieve the Genie ML performance in these cases for most correlation factors $\rho$. Its superiority is hence not only limited to the single case of $\rho=1$, but is present throughout the whole range of $\rho$. For the ChainNet-NN, that has been trained for uncorrelated sources only, we see a quite good performance for low correlations between the individual sources and significant degradation for increasing $\rho$.

	\subsection{Fully Sampled Array}
\begin{figure}[h]
\begin{subfigure}{\columnwidth}
	\centering
	\includegraphics{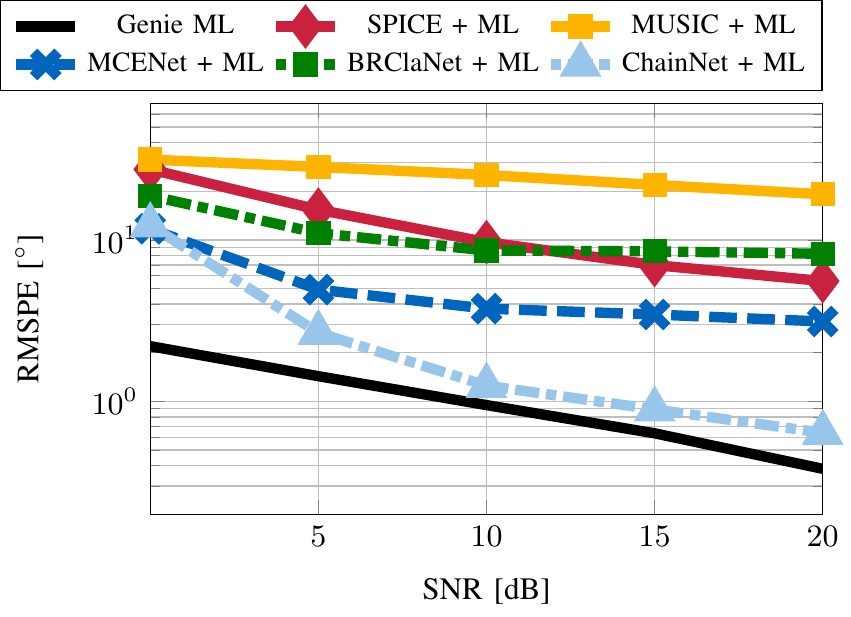}
	\vspace{-5pt}
	
	\caption{RMSPE vs.~SNR for the Top $99\%$ of Realizations}
	\label{fig:rmspeL3full}
\end{subfigure}

\begin{subfigure}{\columnwidth}
	\centering
		\includegraphics{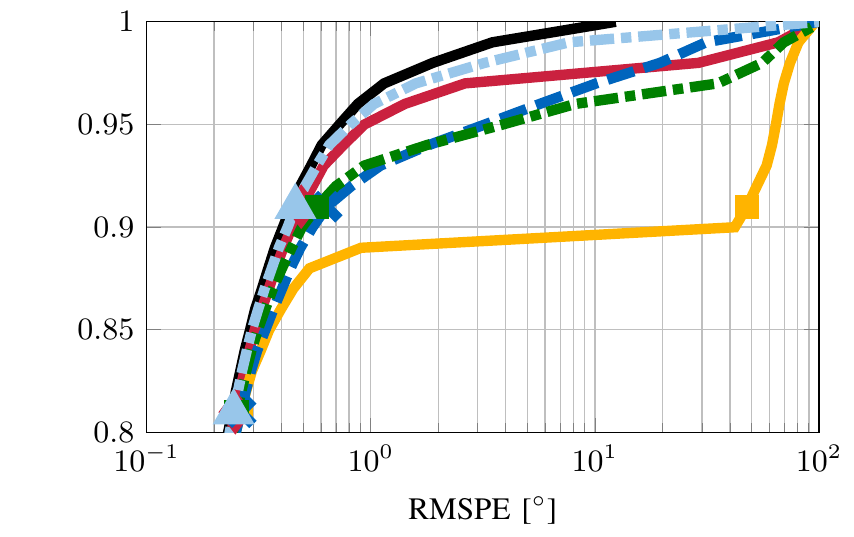}
	\vspace{-5pt}
	
	\caption{Empirical CDF of the fully sampled array at SNR$=20\,\text{dB}$}
	\label{fig:cdfL3full}
\end{subfigure}
\caption{Simulation Results for Fully Sampled Array for $L=3$, $M=9$, $ K=1 $, $W=9$, $N=10$, with gradient-based refinement.}
\end{figure}

\begin{figure}[h]
	\begin{center}
		\includegraphics{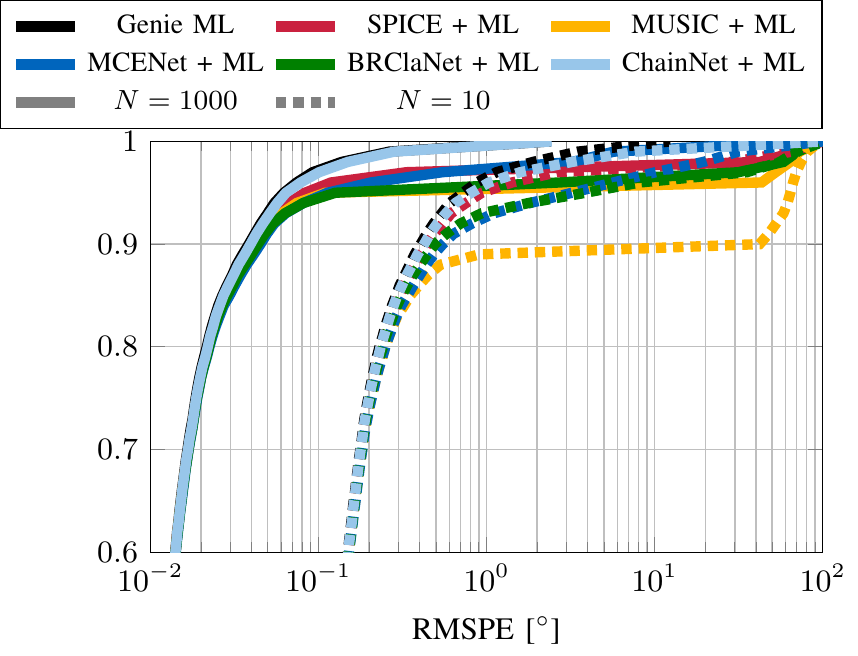}
	\end{center}
	\vspace{-5pt}
	
	\caption{Comparison of Empirical Cumulative Density Functions for Fully Sampled Array for Varying $N$ at SNR$=20\,\text{dB}$ for $L=3$, $M=9$, $ K = 1 $, $ W=9 $, with gradient-based refinement.}
	\label{fig:cdfFullL3N}
\end{figure}

	In the following, we consider a fully sampled array, i.e., $K=1$ and $W=M$. In Figure~\ref{fig:rmspeL3full}, we plot RMSPE of the $99\%$ realizations with the smallest estimation error per DoA estimator. The graphs show that the ChainNet approach does suffer from about $1\%$ of outliers at $10$\,dB and above, whereas the other estimators are less robust. Figure~\ref{fig:cdfL3full} shows the CDF at $20$\,dB SNR for $L=3$ sources. Again, we can observe that the ChainNet-NN comes closest to the Genie ML performance. The SPICE estimator still suffers from fewer outliers than for the system with subarray sampling. In contrast, a plain MUSIC estimator is affected by $10\%$ outliers, which mostly go back to realizations with closely spaced sources. {For a very high number of snapshots (cf. \ref{fig:cdfFullL3N}), the outcome is similar although the MUSIC estimator is now able to produce more meaningful estimates due to the more precise covariance estimate.}
	
	\section{Conclusion}
	In this work, we investigated a new NN-based successive DoA estimator that is superior to existing NN-based and conventional estimators. The proposed ChainNet-NN does not only achieve a very high robustness towards outliers, but also attains a very high accuracy for non-outlier realizations with and without consecutive gradient steps on the likelihood function. Moreover, the presented estimation method provides a nice interpretation as an approximation of a MAP estimator.
	
	Future studies might investigate if passing the complete output of each binary classifier of a certain node to the next node, instead of only the associated DoA estimate, may improve the performance even further. However, note that by passing this soft information between nodes, we can no longer train multiple nodes simultaneously under the assumption of perfect estimates in the previous nodes.
	
	% -------------------------------------------------------------------------
	\bibliographystyle{IEEEtran}
	\bibliography{DoALitDB}

\begin{thebibliography}{10}
\providecommand{\url}[1]{#1}
\csname url@samestyle\endcsname
\providecommand{\newblock}{\relax}
\providecommand{\bibinfo}[2]{#2}
\providecommand{\BIBentrySTDinterwordspacing}{\spaceskip=0pt\relax}
\providecommand{\BIBentryALTinterwordstretchfactor}{4}
\providecommand{\BIBentryALTinterwordspacing}{\spaceskip=\fontdimen2\font plus
\BIBentryALTinterwordstretchfactor\fontdimen3\font minus
  \fontdimen4\font\relax}
\providecommand{\BIBforeignlanguage}[2]{{%
\expandafter\ifx\csname l@#1\endcsname\relax
\typeout{** WARNING: IEEEtran.bst: No hyphenation pattern has been}%
\typeout{** loaded for the language `#1'. Using the pattern for}%
\typeout{** the default language instead.}%
\else
\language=\csname l@#1\endcsname
\fi
#2}}
\providecommand{\BIBdecl}{\relax}
\BIBdecl

\bibitem{Krim1996}
H.~{Krim} and M.~{Viberg}, ``Two decades of array signal processing research:
  The parametric approach,'' \emph{IEEE Signal Process. Mag.}, vol.~13, no.~4,
  pp. 67--94, Jul. 1996.

\bibitem{Trees2002}
H.~L.~V. Trees, \emph{Optimum Array Processing: Part {IV} of Detection,
  Estimation, and Modulation Theory}.\hskip 1em plus 0.5em minus 0.4em\relax
  Wiley, 2002.

\bibitem{Barthelme2020b}
A.~Barthelme and W.~Utschick, ``A machine learning approach to {DoA} estimation
  and model order selection for antenna arrays with subarray sampling,''
  \emph{accepted for publication in IEEE Trans. Sig. Process.}, pp. 1--13, May
  2021.

\bibitem{Chakrabarty2017}
S.~Chakrabarty and E.~A.~P. Habets, ``Broadband {DOA} estimation using
  convolutional neural networks trained with noise signals,'' \emph{Proc.
  WASPAA}, pp. 136--140, Oct. 2017.

\bibitem{Liu2018}
Z.-M. Liu, C.~Zhang, and P.~S. Yu, ``Direction-of-arrival estimation based on
  deep neural networks with robustness to array imperfections,'' \emph{IEEE
  Trans. Antennas Propag.}, vol.~66, no.~12, pp. 7315--7327, Dec. 2018.

\bibitem{Chakrabarty2019}
S.~Chakrabarty and E.~A.~P. Habets, ``Multi-speaker {DOA} estimation using deep
  convolutional networks trained with noise signals,'' \emph{IEEE J. Sel.
  Topics Signal Process.}, vol.~13, no.~1, pp. 8--21, Mar. 2019.

\bibitem{Ozanich2020}
E.~Ozanich, P.~Gerstoft, and H.~Niu, ``A feedforward neural network for
  direction-of-arrival estimation,'' \emph{J. Acoustical Soc. Amer.}, vol. 147,
  no.~3, pp. 2035--2048, Mar. 2020.

\bibitem{Yao2020}
Y.~Yao, H.~Lei, and W.~He, ``\BIBforeignlanguage{en}{A-{CRNN}-based method for
  coherent {DOA} estimation with unknown source number},''
  \emph{\BIBforeignlanguage{en}{Sensors}}, vol.~20, no.~8, p. 2296, Jan. 2020.

\bibitem{Papageorgiou2020a}
\BIBentryALTinterwordspacing
G.~K. Papageorgiou, M.~Sellathurai, and Y.~C. Eldar, ``Deep networks for
  direction-of-arrival estimation in low {SNR},'' \emph{arXiv:2011.08848 [cs,
  eess]}, Nov. 2020. [Online]. Available: \url{http://arxiv.org/abs/2011.08848}
\BIBentrySTDinterwordspacing

\bibitem{Wu2019}
L.~Wu, Z.-M. Liu, and Z.-T. Huang, ``Deep convolution network for direction of
  arrival estimation with sparse prior,'' \emph{IEEE Signal Process. Lett.},
  vol.~26, no.~11, pp. 1688--1692, Nov. 2019.

\bibitem{Elbir2020}
A.~M. Elbir, ``{DeepMUSIC}: Multiple signal classification via deep learning,''
  \emph{IEEE Sensors Lett.}, vol.~4, no.~4, pp. 1--4, Apr. 2020.

\bibitem{Izacard2019}
G.~Izacard, B.~Bernstein, and C.~Fernandez-Granda, ``A learning-based framework
  for line-spectra super-resolution,'' in \emph{Proc. ICASSP}, May 2019, pp.
  3632--3636, iSSN: 2379-190X.

\bibitem{Guo2020}
Y.~Guo, Z.~Zhang, Y.~Huang, and P.~Zhang, ``{DOA} estimation method based on
  cascaded neural network for two closely spaced sources,'' \emph{IEEE Signal
  Process. Lett.}, vol.~27, pp. 570--574, Apr. 2020.

\bibitem{Bialer2019}
O.~Bialer, N.~Garnett, and T.~Tirer, ``Performance advantages of deep neural
  networks for angle of arrival estimation,'' in \emph{Proc. ICASSP}, May 2019,
  pp. 3907--3911.

\bibitem{Barthelme2021}
A.~Barthelme and W.~Utschick, ``{DoA} estimation using neural network-based
  covariance matrix reconstruction,'' \emph{IEEE Signal Processing Letters},
  vol.~28, pp. 783--787, 2021.

\bibitem{Schmidt1986}
R.~Schmidt, ``Multiple emitter location and signal parameter estimation,''
  \emph{IEEE Trans. Antennas Propag.}, vol.~34, no.~3, pp. 276--280, Mar. 1986.

\bibitem{Suleiman2018}
W.~Suleiman, P.~Parvazi, M.~Pesavento, and A.~M. Zoubir, ``Non-coherent
  direction-of-arrival estimation using partly calibrated arrays,'' \emph{IEEE
  Trans. Signal Process.}, vol.~66, no.~21, pp. 5776--5788, Nov. 2018.

\bibitem{Li2018}
Q.~Li, X.~Zhang, and H.~Li, ``Online direction of arrival estimation based on
  deep learning,'' in \emph{Proc. ICASSP}, Apr. 2018, pp. 2616--2620.

\bibitem{Bridle1989}
J.~S. Bridle, ``Training stochastic model recognition algorithms as networks
  can lead to maximum mutual information estimation of parameters,'' in
  \emph{Advances Neural Inf. Process. Syst.}, Nov. 1989, pp. 211--217.

\bibitem{Lecun1998}
Y.~Lecun, L.~Bottou, Y.~Bengio, and P.~Haffner, ``Gradient-based learning
  applied to document recognition,'' \emph{Proc. IEEE}, vol.~86, pp.
  2278--2324, Nov. 1998.

\bibitem{Zhang2014}
M.~Zhang and Z.~Zhou, ``A review on multi-label learning algorithms,''
  \emph{IEEE Trans. Knowl. Data Eng.}, vol.~26, no.~8, pp. 1819--1837, Aug.
  2014.

\bibitem{Herrera2016}
F.~Herrera, F.~Charte, A.~J. Rivera, and M.~J.~d. Jesus,
  \emph{\BIBforeignlanguage{en}{Multilabel Classification: Problem Analysis,
  Metrics and Techniques}}.\hskip 1em plus 0.5em minus 0.4em\relax Springer
  International Publishing, 2016.

\bibitem{Boutell2004}
M.~R. Boutell, J.~Luo, X.~Shen, and C.~M. Brown,
  ``\BIBforeignlanguage{en}{Learning multi-label scene classification},''
  \emph{\BIBforeignlanguage{en}{Pattern Recognit.}}, vol.~37, no.~9, pp.
  1757--1771, Sep. 2004.

\bibitem{Read2011}
J.~Read, B.~Pfahringer, G.~Holmes, and E.~Frank,
  ``\BIBforeignlanguage{en}{Classifier chains for multi-label
  classification},'' \emph{\BIBforeignlanguage{en}{Mach. Learn.}}, vol.~85,
  no.~3, p. 333, Jun. 2011.

\bibitem{Read2020}
------, ``Classifier chains: A review and perspectives,''
  \emph{arXiv:1912.13405 [cs, stat]}, Apr. 2020, arXiv: 1912.13405.

\bibitem{Tensorflow2015}
M.~Abadi, A.~Agarwal, P.~Barham, E.~Brevdo, Z.~Chen, C.~Citro, G.~S. Corrado,
  A.~Davis, J.~Dean, M.~Devin, S.~Ghemawat, I.~Goodfellow, A.~Harp, G.~Irving,
  M.~Isard, Y.~Jia, R.~Jozefowicz, L.~Kaiser, M.~Kudlur, J.~Levenberg,
  D.~Man\'{e}, R.~Monga, S.~Moore, D.~Murray, C.~Olah, M.~Schuster, J.~Shlens,
  B.~Steiner, I.~Sutskever, K.~Talwar, P.~Tucker, V.~Vanhoucke, V.~Vasudevan,
  F.~Vi\'{e}gas, O.~Vinyals, P.~Warden, M.~Wattenberg, M.~Wicke, Y.~Yu, and
  X.~Zheng, ``{TensorFlow}: Large-scale machine learning on heterogeneous
  systems,'' 2015, software available from tensorflow.org.

\bibitem{Costa1999}
P.~{Costa}, J.~{Grouffaud}, P.~{Larzabal}, and H.~{Clergeot}, ``Estimation of
  the number of signals from features of the covariance matrix: A supervised
  approach,'' \emph{IEEE Trans. Signal Process.}, vol.~47, no.~11, pp.
  3108--3115, Nov. 1999.

\bibitem{Niu2017}
H.~Niu, P.~Gerstoft, and E.~Reeves, ``Source localization in an ocean waveguide
  using supervised machine learning,'' \emph{J. Acousti. Soc. Amer.}, vol. 142,
  no.~3, pp. 1176--1188, Sep. 2017.

\bibitem{Barthelme2020}
A.~Barthelme, R.~Wiesmayr, and W.~Utschick, ``Model order selection in {DoA}
  scenarios via cross-entropy based machine learning techniques,'' in
  \emph{Proc. ICASSP}, May 2020, pp. 4622--4626.

\bibitem{Cybenko1989}
G.~Cybenko, ``\BIBforeignlanguage{en}{Approximation by superpositions of a
  sigmoidal function},'' \emph{\BIBforeignlanguage{en}{Math. of Control,
  Signals and Syst.}}, vol.~2, no.~4, pp. 303--314, Dec. 1989.

\bibitem{Hornik1991}
K.~Hornik, ``\BIBforeignlanguage{en}{Approximation capabilities of multilayer
  feedforward networks},'' \emph{\BIBforeignlanguage{en}{Neural Netw.}},
  vol.~4, no.~2, pp. 251--257, Jan. 1991.

\bibitem{Stoica2004}
P.~{Stoica} and Y.~{Selen}, ``Model-order selection: A review of information
  criterion rules,'' \emph{IEEE Signal Process. Mag.}, vol.~21, no.~4, pp.
  36--47, Jul. 2004.

\bibitem{Fleury99}
B.~Fleury, M.~Tschudin, R.~Heddergott, D.~Dahlhaus, and K.~Ingeman~Pedersen,
  ``Channel parameter estimation in mobile radio environments using the {SAGE}
  algorithm,'' \emph{IEEE Journal on Selected Areas in Communications},
  vol.~17, no.~3, pp. 434--450, 1999.

\bibitem{Stoica2011}
P.~Stoica, P.~Babu, and J.~Li, ``{SPICE}: A sparse covariance-based estimation
  method for array processing,'' \emph{IEEE Trans. Signal Process.}, vol.~59,
  no.~2, pp. 629--638, Feb. 2011.

\bibitem{Glorot2010}
X.~Glorot and Y.~Bengio, ``Understanding the difficulty of training deep
  feedforward neural networks,'' in \emph{Proc. Int. Conf. Artif. Intell.
  Statist.}, vol.~9, May 2010, pp. 249--256.

\bibitem{Kingma2014}
\BIBentryALTinterwordspacing
D.~P. Kingma and J.~Ba, ``Adam: A method for stochastic optimization,''
  \emph{arXiv:1412.6980 [cs.LG]}, Dec. 2014. [Online]. Available:
  \url{https://arxiv.org/abs/1412.6980}
\BIBentrySTDinterwordspacing

\bibitem{Sheinvald1999}
J.~Sheinvald and M.~Wax, ``Direction finding with fewer receivers via
  time-varying preprocessing,'' \emph{IEEE Trans. Signal Process.}, vol.~47,
  no.~1, pp. 2--9, Jan. 1999.

\end{thebibliography}
	
\end{document}